**Response to I. I. Mazin's correspondence on "Electronic correlations in the iron pnictides"**

M. M. Qazilbash[1], J. J. Hamlin[1,2], R. E. Baumbach[1,2], Lijun Zhang[3], D. J. Singh[3], M. B. Maple[1,2] and D. N. Basov[1].

1. Department of Physics, University of California, San Diego, La Jolla, California 92093, USA
2. Institute for Pure and Applied Physical Sciences, University of California, San Diego, La Jolla, California 92093, USA
3. Materials Science and Technology Division, Oak Ridge National Laboratory, Oak Ridge, Tennessee 37831, USA

There is accumulating evidence for substantial renormalization of the iron *d*-bands on a several tenth of eV scale in the iron pnictides, starting with photoemission experiments on LaFePO [1]. Recent direct measurements of the Fermi surface of LaFePO using quantum oscillations support this, finding mass renormalizations of ~2 depending on the particular Fermi surface sheet compared with band structure calculations [2]. We compared optical spectra and electronic structure calculations and discovered a renormalization of the Drude weight in LaFePO consistent with the mass renormalization found in the above experiments [3]. Specifically, we showed that the experimental Drude weight ($K_{exp}$) is substantially suppressed relative to the bare band structure value ($K_{band}$) and we attributed this phenomenon to correlation effects of electronic origin [3].

Recently, Mazin has criticized our conclusion arguing that band shifts of +100 meV for the electron bands and -100 meV for the hole bands would reduce the needed mass renormalization by ~2/3, leaving only the remaining ~1/3 to be explained by electronic correlations [4]. In this response, we point out that the effect discussed by Mazin is considerably smaller than he claims and, in particular, is too small to change our conclusions. In addition, the minor band shifts may well be a consequence of the electronic correlations that we find.

Mazin bases his criticism on the measurements by Coldea and co-workers [2]. Actually, Coldea and co-workers note that shifts will improve the agreement between the density functional bands and their quantum oscillation measurements, but the shifts are substantially smaller than those assumed by Mazin, and even then some of the shifts may be related to sample stoichiometry. None of these shifts on any Fermi surface is as large as 100 meV quoted by Mazin. A noteworthy effect is the removal of the heavy 3D hole sheet. However, because of its heavy mass, this sheet contributes very little to the Drude weight that we register in our experiments. The main contribution (~65%) to the Drude weight according to our calculations in fact comes from the electron sections. This dominance of the electron sheets is important because the overall effect of band shifts on the electron contribution are smaller than the effect of similar shifts on the hole contribution. There are two electron sections, an inner section and an outer section, which according to Coldea and co-workers can be brought into optimum agreement with their experiment using shifts of 83 and 30 meV accompanied by a mass renormalization. The optimum shift of the hole bands of -53 meV is also much smaller in magnitude than -100 meV invoked by Mazin. The actual change in the electron Fermi surface volume is small (~15%), as may be seen from Figs. 3c and 3d of Ref. 2. This contradicts the claim of Mazin, who asserts that the main contribution to the discrepancy in the optical Drude spectral weight (proportional to $n/m_{opt}$) is due to a change in the effective Fermi surface volume ($n$) and not the effective optical mass ($m_{opt}$).

Finally, we note that although Mazin invokes the well known density functional band gap error in simple insulators such as Ge as the mechanism for band shifts, there is no reason to suppose that the same physics is operative in a partially filled metallic $d$-band system. For example, there are strong shifts in electronic structure relative to density functional calculations in correlated oxides such as NiO, and these are primarily associated with electron correlations (Mott physics in that case).

To summarize, Mazin's arguments do not invalidate our conclusions because (1) the band shifts seen in experiment are significantly smaller than what he proposed, leaving a substantial renormalization of the Drude weight to be explained by correlations of electronic origin, (2) the observed band shifts may also reflect electronic correlations themselves and (3) many experiments are converging on the fact that there are substantial band renormalizations on the tenths of eV scale consistent with our report. These include optical measurements on $EuFe_2As_2$ as compared with density functional calculations by

Moon, Mazin and co-workers [5]. Moreover, our conclusion about the importance of electronic correlations in the iron pnictides was also based on the significant renormalization of the Drude weight ($K_{exp}/K_{band} \sim 0.3$) in paramagnetic $BaFe_2As_2$ [3]. A more recent infrared work on LaFeAsO also reported a factor-of-3 renormalization of the Drude weight [6]. As for the detailed nature of the electronic correlation effects, this remains an open question [3].

[1]     D. H. Lu *et al.*, Nature **455**, 81 (2008).
[2]     A. I. Coldea *et al.*, Phys. Rev. Lett. **101**, 216402 (2008).
[3]     M. M. Qazilbash *et al.*, Nature Physics **5**, 647 (2009).
[4]     I. I. Mazin, arXiv:0910.4117 (2009).
[5]     S. J. Moon *et al.*, arXiv:0909.3352 (2009).
[6]     Z. G. Chen *et al.*, arXiv:0910.1318 (2009).